# Differential diffusion effects and super-adiabatic local temperature in lean hydrogen-air turbulent flames

H.C. Lee[a], P. Dai[a,*], M. Wan[a,b], Andrei N. Lipatnikov[c]

[a]*Guangdong Provincial Key Laboratory of Turbulence Research and Applications, Department of Mechanics and Aerospace Engineering, Southern University of Science and Technology, Shenzhen 518055, People's Republic of China.*
[b]*Jiaxing Research Institute, Southern University of Science and Technology, Jiaxing, 314031, Zhejiang, People's Republic of China.*
[c]*Department of Mechanics and Maritime Sciences, Chalmers University of Technology, Gothenburg SE-412 96, Sweden*

**Abstract**
Analyzed in this paper are three-dimensional Direct Numerical Simulation (DNS) data obtained from seven statistically planar and one-dimensional, lean complex-chemistry hydrogen-air flames propagating in a box with forced turbulence. The simulation conditions cover a wide range of non-dimensional turbulent combustion characteristics. Specifically, root-mean-square turbulent velocity is varied from 2.2 to 54 laminar flame speeds, integral length scale of turbulence is varied from 0.5 to 2.2 laminar flame thicknesses, Damköhler and Karlovitz number are varied from 0.01 to 0.53 and from 10 to 1315, respectively. Two equivalence ratios, 0.5 and 0.35, are explored. Turbulent burning velocities are evaluated for these seven low Lewis number flames and equidiffusion counterparts to six of them. Moreover, conditioned profiles of temperature, fuel consumption and heat release rates and probabilities of finding superadiabatic temperature are sampled from all seven low Lewis number flames. Analyses of obtained results show that both magnitude of superadiabatic temperature and probability of finding it are decreased with increasing Karlovitz number $Ka$. However, significant influence of differential diffusion effects on local structure of flame reaction zones and bulk burning velocity is well pronounced in all cases, even at $Ka$ as high as 1315. Therefore, a decrease in magnitude of superadiabatic local temperature with increasing Karlovitz number or even negligible probability of finding such a high temperature at high $Ka$ is not an evidence that differential diffusion effects play a minor role under such conditions. The simulated mitigation of phenomenon of superadiabatic temperature at high $Ka$ is attributed to intensification of turbulent mixing in local flame oxidation zones, rather than weakening differential diffusion effects in local flame reaction zones.

## 1. Introduction

As reviewed elsewhere [1-5], turbulent burning of lean $H_2$-containing mixtures is greatly affected by differences between molecular diffusion coefficients of hydrogen, oxygen, and heat. On the global level, such effects manifest themselves in abnormally high ratios of turbulent and laminar burning velocities, revealed in pioneering measurements by Karpov et al. [6,7] and in many subsequent experiments, see review articles [2,4] and Refs. [8-14] as recent examples. On the local level, these differential diffusion effects manifest themselves in significant variations in the equivalence ratio $\phi$, species mass fractions $Y_k$, and temperature $T$, including appearance of hot spots characterized by super-adiabatic temperatures. Such local phenomena were recently explored in complex-chemistry Direct Numerical Simulation (DNS) studies [15-27] and experiments [28,29]. These phenomena (i) are commonly attributed to imbalance of molecular fluxes of heat and chemical energy from and to, respectively, inherently laminar reaction zones stretched by turbulent eddies and (ii) are widely accepted to result in significantly increasing the normalized turbulent burning rate $U_T/S_L$ [1-5,30]. Here, $S_L$ is the unstretched laminar flame speed.

However, contemporary understanding of the discussed global and local effects does not seem to be self-consistent. On the one hand, the present authors are not aware on an experimental study that shows reduction of the aforementioned increase in $U_T/S_L$ (caused by differential diffusion effects) with increasing the normalized rms velocity $u'/S_L$ or/and Karlovitz number $Ka \propto (u'/S_L)^{3/2}(L/\delta_L)^{-1/2}$. Here, $L$ is an integral length scale of turbulence and $\delta_L$ is a laminar flame thickness. Abnormally large ratios of $U_T/S_L$ were well documented in experiments at extremely high $Ka \gg 1$ [6-8] or at smaller but still very large $Ka = O(10^3)$ [12,13]. Such data imply that even very intense turbulence does not suppress differential diffusion effects on the global level.

On the other hand, both DNS [16-18,26,27] and experimental [28,29] data indicate reduction of local variations in the equivalence ratio or/and magnitude of super-adiabatic temperature with increasing $Ka$, thus, implying mitigation of differential diffusion effects on the local level. In particular, a decrease in the peak (super-adiabatic) local temperature with increasing $Ka$ is often considered to indicate decreasing importance of differential diffusion effects in highly turbulent flames [5,26,27]. However, simultaneous survival and disappearance of such effects on the global and local levels, respectively, does not seem to be self-consistent. Accordingly, the present work aims at exploring whether or not (i) an increase in $U_T/S_L$ due to differential diffusion effects and (ii) a decrease in magnitude of super-adiabatic temperature with increasing $Ka$ are consistent with one another in highly turbulent premixed flames.

For this purpose, a simple hypothesis is put forward in Sect. 2. Subsequently, the hypothesis is assessed by analyzing previous and new DNS data described in Sect. 3. Results are reported and discussed in Sect. 4, followed by conclusions.

## 2. A simple hypothesis

A typical complex-chemistry laminar flame is well known to involve a preheat zone, an inner layer, and an oxidation layer [31,32]. Widely recognized combustion regime diagrams [33-36] and most models of premixed turbulent combustion deal with major characteristics of preheat zone and inner layer, e.g., their thicknesses, which are very different (the former zone is commonly considered to be significantly thicker [35]). Processes localized to oxidation layer are beyond the scope of mainstream approaches to modeling the influence of turbulence on premixed combustion, reviewed elsewhere [3,37-40]. On the contrary the focus of the present study is placed on phenomena localized to inner and oxidation layers.

As discussed in detail elsewhere [31,32,35], burning rate is controlled by processes localized to a thin inner layer. Therein, fuel consumption and heat release rates reach their peak values, but temperature is substantially lower than its adiabatic value $T_{ad}$. In a significantly thicker oxidation layer, the temperature continues growing due to heat release in slow termolecular radical recombination reactions, but the local heat release rate is significantly lower than its peak value and the fuel consumption rate almost vanishes. Thus, processes localized to an oxidation layer substantially affect temperature field, contribute weakly to heat release, and do not affect fuel consumption. Moreover, due to the relatively large thickness of an oxidation layer, small turbulent eddies (that are sufficiently strong to survive in the local flames) could enter the layer and accelerate mixing within it.

Accordingly, the following physical scenario is worth considering. First, even at high $Ka$, differential diffusion can greatly affect the inner layer structure and result locally in increasing temperature, enthalpy, fuel consumption rate, etc. (at least, at the leading edge of the flame brush [19]). Therefore, $U_T/S_L$ is also increased due to differential diffusion, as evidenced in numerous experiments [6-8,12,13,41]. Second, when a fluid volume with the local enthalpy excess (arisen due to differential diffusion) moves through the oxidation layer, the volume temperature (i) is increased due to heat release in termolecular recombination reactions, but (ii) is decreased due to small-scale-turbulence mixing with surrounding volumes characterized by lower enthalpies (weaker affected by differential diffusion). At low or moderate $u'/S_L$ and $Ka$, the former process dominates, the local temperature continues growing and can exceed $T_{ad}$. In other words, small-scale turbulent mixing is too weak to smooth out potential hot spots originating within the inner layer due to differential diffusion. When $u'/S_L$ and $Ka$ are increased, the local turbulent mixing is intensified within the oxidation layer and mitigates the local temperature increase. At even higher $u'/S_L$ and $Ka$, the mixing could overwhelm the local heat release, and the local temperature could remain smaller than $T_{ad}$.

This physical scenario admits both (i) a significant increase in $U_T/S_L$ due to the influence of differential diffusion on thin-inner-layer structure and (ii) the lack of super-adiabatic temperatures due to intense small-scale-turbulence mixing within a thicker oxidation layer. In the rest of the paper, this scenario is assessed by analyzing DNS data.

## 3. DNS attributes and diagnostic methods

Since most of the DNS data analyzed in the following were already reported in our previous papers [19, 42-48], we will restrict ourselves to a summary of the major DNS attributes and refer the interested reader to the cited papers.

This study addresses statistically planar and one-dimensional, lean $H_2$-air, complex-chemistry flames propagating in a box of $\Lambda \times M\Lambda \times \Lambda$ in forced turbulence along $y$-direction. Numerical meshes consist of $N \times MN \times N$ cells. The length $\Lambda$ and the numbers $M$ and $N$ are reported in Table 1. Inflow and outflow boundary conditions are set along the streamwise direction $y$. Other boundary conditions are periodic.

The flames were simulated adopting software *DINO* [49], which numerically solved unsteady and three-dimensional continuity, Navier-Stokes, energy, and species transport equations written in the low-Mach-number formulation. The ideal gas state equation, a detailed chemical mechanism by Kéromnès et al. [50], and mixture-averaged molecular transport model [51] were also invoked.

To generate turbulence, variable-density version [52,53] of a linear forcing method [54-56] was used. Combustion simulations were started by embedding a planar laminar flame (computed using *Cantera* [51]) in the center of the computational domain at $t = 0$.

Table 1.
Major characteristics of simulated flames.

| Case | $\phi$ | $S_L^0$, m/s | $\delta_L^T$, mm | Le | $\frac{u'}{S_L^0}$ | $\frac{L}{\delta_L^T}$ | Da | Ka | $Re_\lambda$ | $\frac{\Delta x}{\eta_K}$ | N | M | $\Lambda$, mm | $\frac{\bar{\bar{U}}_T}{S_L^0}$ |
|---|---|---|---|---|---|---|---|---|---|---|---|---|---|---|
| A | 0.5 | 0.58 | 0.41 | 0.39 | 2.2 | 1.1 | 0.53 | 9.9 | 21 | 1.08 | 64 | 18 | 2.4 | 4.1 |
| A1 | 0.5 | 0.78 | 0.29 | 1.0 | 1.6 | 1.6 | 1.0 | 5.4 | 19 | 1.08 | 64 | 18 |  | 2.7 |
| B | 0.5 | 0.58 | 0.41 | 0.39 | 4.0 | 1.1 | 0.29 | 26.0 | 26 | 1.13 | 96 | 18 | 2.4 | 5.4 |
| B1 | 0.5 | 0.78 | 0.29 | 1.0 | 3.0 | 1.6 | 0.54 | 13.9 | 26 | 1.13 | 96 | 18 |  | 3.2 |
| C | 0.5 | 0.58 | 0.41 | 0.39 | 11.2 | 1.1 | 0.10 | 121. | 44 | 1.85 | 128 | 16 | 2.4 | 8.2 |
| C1 | 0.5 | 0.78 | 0.29 | 1.0 | 8.3 | 1.6 | 0.19 | 66.6 | 43 | 1.85 | 128 | 16 |  | 4.3 |
| $C_{Da}$ | 0.5 | 0.58 | 0.41 | 0.39 | 22.4 | 2.2 | 0.10 | 246. | 92 | 1.75 | 384 | 7 | 4.8 | 9.8 |
| D | 0.35 | 0.12 | 0.92 | 0.36 | 54.1 | 0.5 | 0.01 | 1315. | 44 | 1.85 | 128 | 16 | 2.4 | 14.9 |
| D1 | 0.35 | 0.30 | 0.43 | 1.0 | 21.6 | 1.1 | 0.05 | 268. | 41 | 1.85 | 128 | 16 |  | 3.4 |
| E | 0.35 | 0.12 | 0.92 | 0.36 | 11.2 | 0.5 | 0.04 | 125. | 20 | 1.13 | 64 | 16 | 2.4 | 12.6 |
| E1 | 0.35 | 0.30 | 0.43 | 1.0 | 4.5 | 1.1 | 0.24 | 24.3 | 20 | 1.13 | 64 | 16 |  | 2.2 |
| F | 0.35 | 0.12 | 0.92 | 0.36 | 11.2 | 1.2 | 0.10 | 85.7 | 29 | 1.07 | 128 | 16 | 5.6 | 22.4 |
| F1 | 0.35 | 0.30 | 0.43 | 1.0 | 4.5 | 2.5 | 0.56 | 85.7 | 28 | 1.07 | 128 | 16 |  | 4.5 |

Major characteristics of the studied flames are reported in Table 1, where the laminar flame speed $S_L^0$ has been computed using the same chemical mechanism [50] and software *Cantera* [51]; $\delta_L^T = (T_b - T_u)/\max|\nabla T|$ and $\tau_F = \delta_L^T/S_L^0$ designate the laminar flame thickness and time scale, respectively; $u'$ and $L = u'^3/\varepsilon_0$ are target values of rms velocity and integral length scale of turbulence, respectively; $\varepsilon_0 = 2\nu\langle S_{ij}S_{ij}\rangle$ is dissipation rate averaged over forced turbulence volume before embedding a flame; $S_{ij} = (\partial u_i/\partial x_j + \partial u_j/\partial x_i)/2$ is the rate-of-strain tensor; $\nu$ is kinematic viscosity of unburned reactants; and the summation convention applies to repeated indexes. The linear forcing method yields $L = 0.19\Lambda$ [41,52,53]; Karlovitz number $Ka = \tau_F/\tau_K$, Reynolds number $Re_\lambda = u'\lambda/\nu$, and Damköhler number $Da = \tau_t/\tau_F$ have been evaluated using Kolmogorov time scale $\tau_K = (\nu_u/\varepsilon_{le})^{1/2}$, Taylor microscale $\lambda = (10\nu_u u'/\varepsilon_{le})^{1/2}$, and the integral time scale $\tau_t = L/u'$, respectively; the dissipation rate $\varepsilon_{le}$ is averaged over flame-brush leading zone where transverse-averaged combustion progress variable $0.01 < \bar{c}_F(y,t) < 0.05$. It is defined adopting the fuel mass fraction $Y_F$, i.e., $c_F = 1 - Y_F/Y_{F,u}$. Finally, $\eta_K = (\nu^3/\varepsilon_{le})^{1/4}$ is Kolmogorov length scale;

$\Delta x$ is the mesh step; and Lewis number $Le = a_u/D_{H_2,u}$ is equal to a ratio of heat and hydrogen diffusivities in unburned gas.

Cases A-F and A1-F1 were simulated earlier [19, 42-48]. Number 1 in the case names refers to equidiffusive flames, which were explored by setting $D_k = a$ for all species. Case C$_{Da}$ is a new case. It is designed based on case C by retaining the same $Da$, but increasing the computational domain width $\Lambda$ and $u'$ to increase $Ka$ by a factor of two. This new case was designed for another study, but is also analyzed here to expand the range of variations in $Ka$ at $\phi = 0.5$. Thus, there are two groups of low Lewis number flames; (i) four lean ($\phi = 0.5$) flames A, B, C, and C$_{Da}$, characterized by different Karlovitz numbers, which are increased from case A to case C$_{Da}$, and (ii) three very lean ($\phi = 0.35$) flames D, E, and F, characterized by different $Ka$ decreased from case D to case F.

Reported in the last column in Table 1 is the mean normalized burning velocities evaluated by spatially integrating the mass Fuel Consumption Rate (FCR), i.e.,

$$U_T(t) = -\frac{W_F}{\rho_u Y_{F,u} \Lambda^2} \iiint \dot{\omega}_F(\mathbf{x}, t) d\mathbf{x}, \quad (1)$$

followed by time-averaging. Here, $\rho$ is the density and $W_F$ is fuel molecular weight. For flames A-F and their equidiffusive counterparts A1-F1, these results were discussed in detail in earlier papers [42-45]. For the goals of the present study, it is worth stressing that differential diffusion effects are well pronounced even in case D characterized by $Ka$ as high as 1315. Indeed, in that flame, $\overline{U_T(t)}/S_L^0$ is larger than its equidiffusive counterpart (case D1) by a factor of 4.4. Nevertheless, a smaller ratio of $(\overline{U_T}/S_L^0)_D/(\overline{U_T}/S_L^0)_{D1} = 4.4$ when compared to $(\overline{U_T}/S_L^0)_E/(\overline{U_T}/S_L^0)_{E1} = 5.7$ implies that the effect magnitude is decreased when $Ka$ is increased from 125 (case E) to 1315 (case D). Note also that $(\overline{U_T}/S_L^0)_F/(\overline{U_T}/S_L^0)_{F1} = 5.0$, i.e., the effect magnitude is larger at $Ka = 125$ (case E) when compared to $Ka = 86$ (case F).

The following discussion is based on analyses of the conditioned profiles $\langle T|c_F\rangle(c_F)$, $\langle\dot{\omega}_F|c_F\rangle(c_F)$, and $\langle\dot{\omega}_T|c_F\rangle(c_F)$ of temperature, FCR, and Heat Release Rate (HRR), respectively, obtained from different low-Lewis-number flames. Methods adopted to sample such time-averaged conditioned profiles have already been described in detail [19]. Moreover, from each bin of $c_F(\mathbf{x}, t)$, the peak values of $T(\mathbf{x}, t)$, $\dot{\omega}_F(\mathbf{x}, t)$, and $\dot{\omega}_T(\mathbf{x}, t)$ were also sampled from the entire flame brushes at each instant followed by time-averaging. Besides, probabilities of $T(\mathbf{x}, t) > T_{ad}$ were evaluated either for each bin of $c_F(\mathbf{x}, t)$ or from the entire flame brush by counting the number of points characterized by $T(\mathbf{x}, t) > T_{ad}$ and the total number of points for that bin or flame brush, respectively.

## 4. Results and discussion

Figure 1 presents the normalized conditioned profiles of (a)-(b) FCR and (c)-(d) HRR, i.e., $\langle\dot{\omega}_F|c_F\rangle(c_F)$ and $\langle\dot{\omega}_T|c_F\rangle(c_F)$, respectively. These profiles have been sampled from the entire flame brushes, time-averaged, and normalized using values of the plotted rates, computed at the same $c_F$ in the counterpart unstretched laminar flames, i.e., $\dot{\omega}_{F,L}(c_F)$ and $\dot{\omega}_{T,L}(c_F)$, respectively. Thus, at each $c_F$, all (four or three) profiles shown in a single plot are normalized using the same rate. The profiles are reported at $c_F > 0.4$, because the focus of the present study is placed on processes localized to large $c_F$.

In all cases, the normalized FCR and HRR are larger than unity (with the exception of a zone of $c_F > 0.9$ at $\phi = 0.5$, where the dimensional rates are very low) and are significantly increased with decreasing $c_F$. The computed large magnitudes of FCR and HRR in turbulent flames when compared to unstretched laminar flames are attributed to differential diffusion. They are not further discussed here, because such effects are explored in detail in earlier papers [19,45,47], where the same DNS data obtained from flames A-F and A1-F1 are analyzed. The focus of the present consideration is placed on comparison of the conditioned rates computed in turbulent flames characterized by different $u'/S_L^0$ and by different $Ka$.

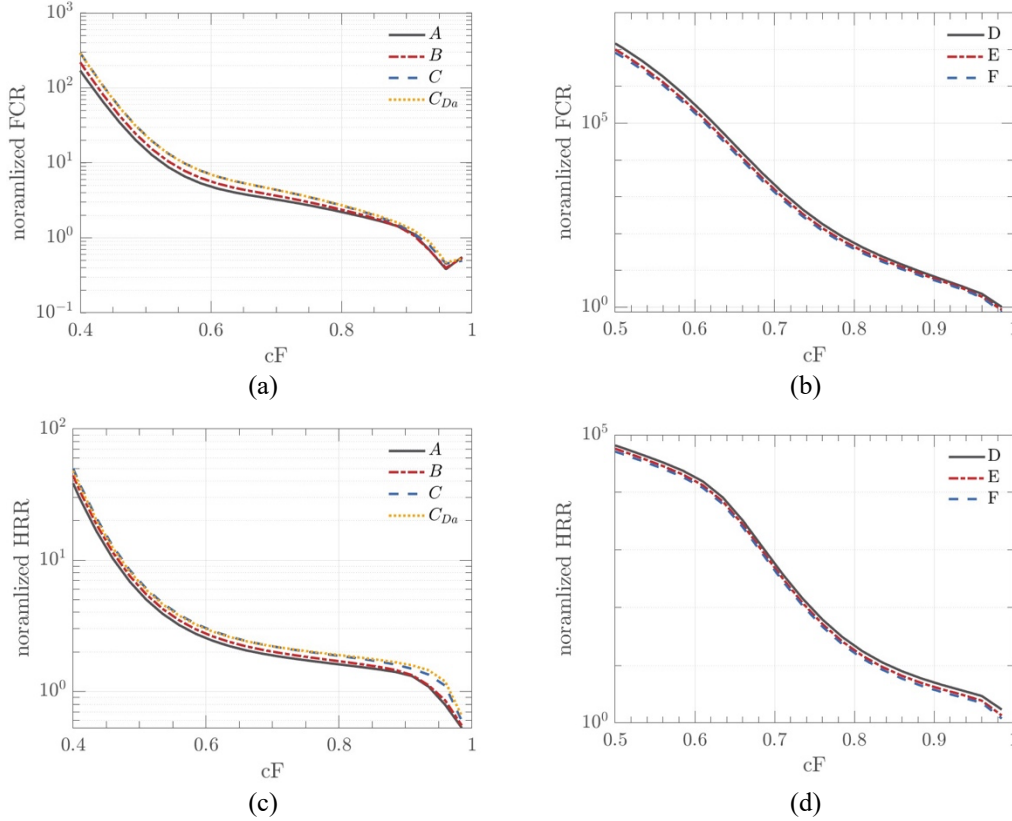

Fig. 1. Normalized time-averaged conditioned profiles of (a)-(b) FCR and (c)-(d) HRR, sampled from (a) and (c) lean ($\phi = 0.5$) flames A-C and $C_{Da}$ or (b) and (d) very lean ($\phi = 0.35$) flames D-F. Profiles are sampled from the entire flame brushes. Normalization is performed using values of the plotted quantities, computed at the same $c_F$ in the unstretched laminar flame.

Figure 1 does not indicate a decrease in the magnitude of the conditioned FCR or HRR with increasing $u'/S_L^0$ and $Ka$ (with the exception of a zone of $c_F > 0.9$ at $\phi = 0.5$). On the contrary, the normalized conditioned rates are increased with increasing $u'/S_L^0$ and $Ka$ (while the effect magnitude appears to be weak, it is worth remembering that the results are presented in logarithmic scale). In lean flames A-C and $C_{Da}$, this trend correlates straightforwardly with an increase in the magnitude of differential effects on the global level. Specifically, ratios of $\overline{U_T(t)}/S_L^0$ obtained from flames A, B, and C and their equidiffusive counterparts A1, B1, and C1, respectively, see the last column in Table 1, are equal to 1.5, 1.7, and 1.9, respectively. In leaner flames, such a straightforward correlation is not observed, e.g., the aforementioned ratio is equal to 4.4, 5.7, or 5.0 for flame pairs D and D1, E and E1, or F and F1, respectively. This lack of correlation could be attributed to the fact that turbulent burning velocity is controlled not only by variations of FCR and HRR within a mean flame brush but also by the flame brush thickness. However, this issue is beyond the scope of the present study. For its major goal, it is sufficient to stress that the last column in Table 1 and Fig. 1 clearly show that significant influence of differential diffusion on burning rate is not mitigated by turbulence on either the global or the local level.

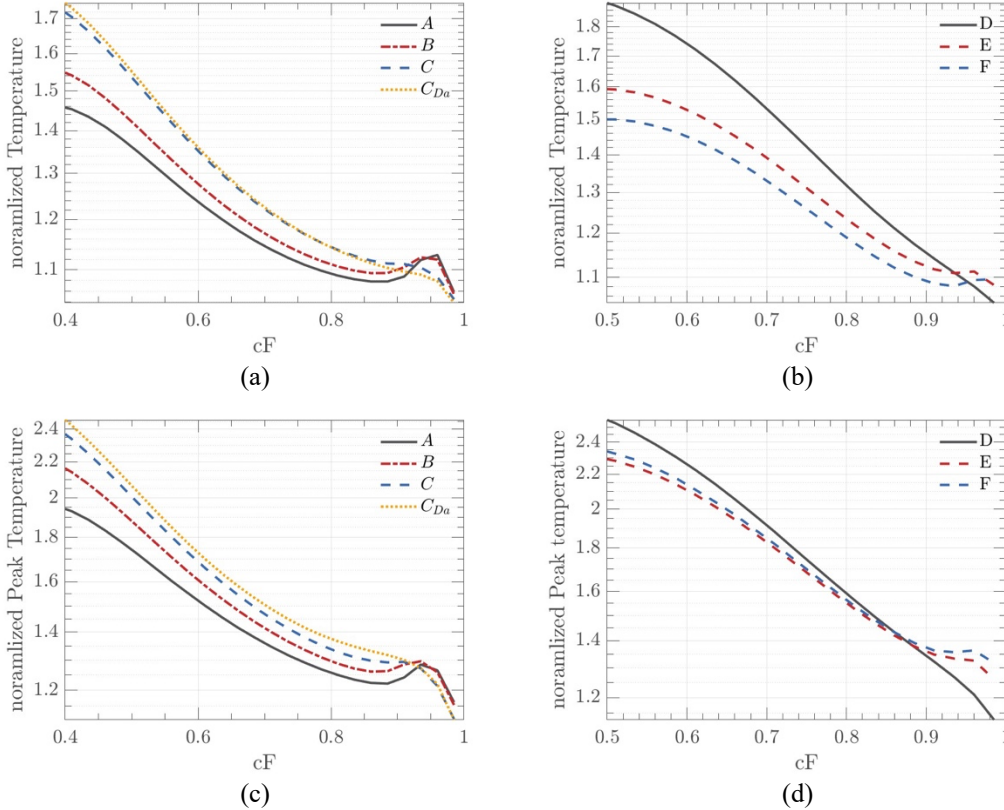

Fig. 2. Profiles of normalized (a)-(b) time-averaged conditioned temperature or (c)-(d) peak local temperature, sampled from (a) and (c) lean ($\phi = 0.5$) flames A-C and $C_{Da}$ or (b) and (d) very lean ($\phi = 0.35$) flames D-F. Profiles are sampled from the entire flame brushes. Normalization is performed using values of the plotted quantities, computed at the same $c_F$ in the unstretched laminar flame.

Figure 2 presents profiles of (a)-(b) the normalized conditioned temperature $\langle T|c_F \rangle(c_F)$ and (c)-(d) the normalized peak local temperature. These profiles have been sampled from the entire flame brushes and normalized using temperatures $T_L(c_F)$, computed at the same $c_F$ in the counterpart unstretched laminar flames. Moreover, the conditioned profiles $\langle T|c_F \rangle(c_F)$ are averaged over time.

Figure 2 shows that an increase in $u'/S_L^0$ and/or $Ka$ results in increasing $\langle T|c_F \rangle(c_F)$ at $c_F < 0.9$, while this effect is weakly pronounced in cases C and $C_{Da}$, where the profiles are close to one another, cf. curves plotted in blue dashed and yellow dotted lines in Fig. 2a. A similar trend is observed for $\max\{T(\mathbf{x},t)\}(c_F)$, with the effect being more (less) pronounced when compared to $\langle T|c_F \rangle(c_F)$ in cases C and $C_{Da}$ (E and F, respectively), cf. curves plotted in blue dashed and yellow dotted lines (blue and dashed red lines, respectively) in Fig. 2a and 2c (Fig. 2b and 2d, respectively). Since cases E and F are characterized by the same $u'/S_L^0$ but different $Ka$, the latter observation could imply that $\max\{T(\mathbf{x},t)\}(c_F)$ is mainly controlled by $u'/S_L^0$, but more cases should be explored to test this simplest interpretation.

For the goals of the present study, the following two trends are of the most importance; (i) the normalized temperatures are larger than unity (due to differential diffusion, as discussed earlier [19,45,47]) and (ii) there is no sign that this effect is mitigated by turbulence if $c_F < 0.9$. Thus, on the local level differential diffusion significantly affects not only FCR and HRR, as shown in Fig. 1, but also temperature in flame reaction zones.

However, such signs are well pronounced at large $c_F$ associated with oxidation layer. Specifically, (i) $\langle T|c_F\rangle(c_F \approx 0.95)$ is decreased with increasing $u'/S_L^0$ and $Ka$ in cases A-C and $C_{Da}$, see Fig. 2a, (ii) $\langle T|c_F\rangle(c_F \approx 0.97)$ is the lowest in case D characterized by the largest $u'/S_L^0$ and $Ka$ at $\phi = 0.35$, see Fig. 2b, (iii) $\max\{T(\mathbf{x},t)\}(c_F > 0.95, \phi = 0.5)$ is lower in cases C and $C_{Da}$, characterized by the two highest $Ka$, when compared to cases A and B, characterized by smaller $Ka$, see Fig. 2c, and (iv) $\max\{T(\mathbf{x},t)\}(c_F > 0.9, \phi = 0.35)$ is decreased with increasing $Ka$, see Fig. 2d.

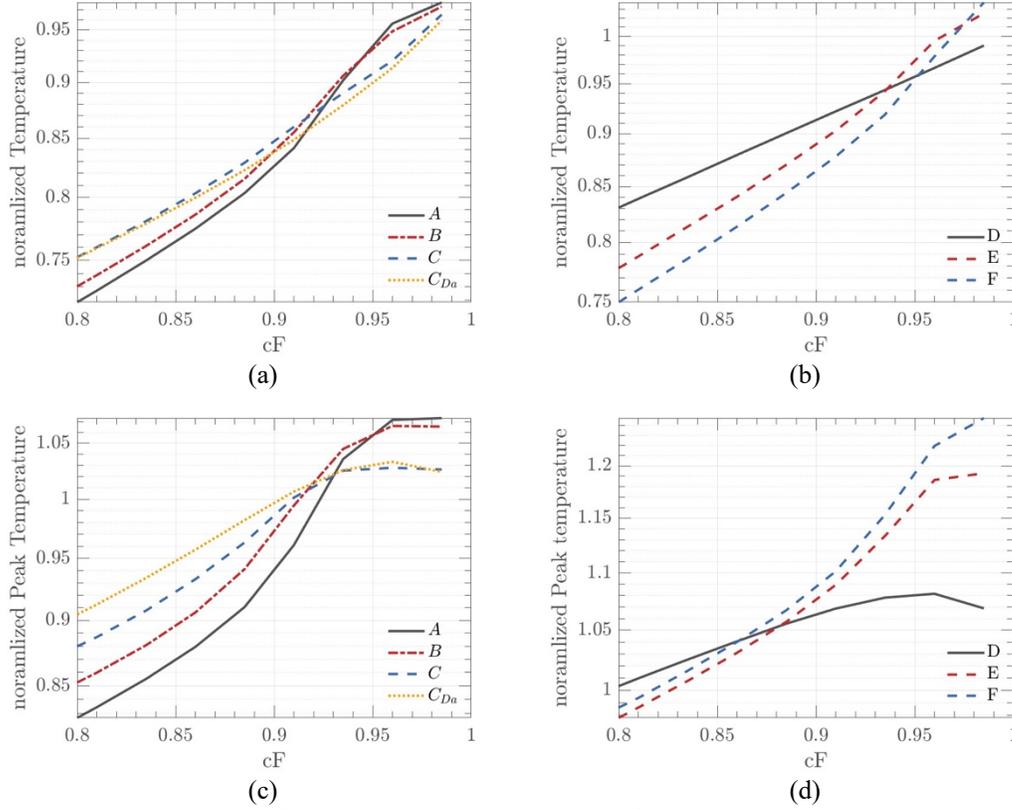

Fig. 3. Temperature profiles reported in Fig. 2 but normalized using $T_{ad}$ and zoomed at large $c_F$. Other details are provided in caption to Fig. 2.

These trends are also observed in Fig. 3, where the same temperature profiles are re-normalized using $T_{ad}$ and are zoomed at large $c_F$. Moreover, Figs. 3c and 3d show that the magnitude of super-adiabatic temperature, which appears at $c_F$ close to unity, is decreased with increasing $Ka$. This trend is not observed in cases C and $C_{Da}$, characterized not only by different $u'/S_L^0$ and different $Ka$, but also by different $L/\delta_L^T$, see Table 1.

Mitigation of a phenomenon of superadiabatic temperature in lean hydrogen-air flames by turbulence is further supported in Fig. 4. At large $c_F$, i.e., if $c_F > 0.945$ and $\phi = 0.5$ or $c_F > 0.88$ and $\phi = 0.35$, Figs. 4a and 4b, respectively, show a decrease in the conditioned probability of finding superadiabatic temperature with increasing $u'/S_L^0$. At each instant, such a probability was evaluated by (i) counting the number of points characterized by $T(\mathbf{x}, t) > T_{ad}$ within a $c_F$-bin and (ii) dividing this number with the total number of grid points associated with the same bin. Subsequently, the results were averaged over time. It is also of interest to note that the reported conditioned probability is weakly affected by variations in $L/\delta_L^T$ at the same $u'/S_L^0$, cf. cases E and F in Fig. 4b. This observation implies that appearance of hot spots characterized by $T(\mathbf{x}, t) > T_{ad}$ is a small-scale phenomenon, but further research into this issue is required.

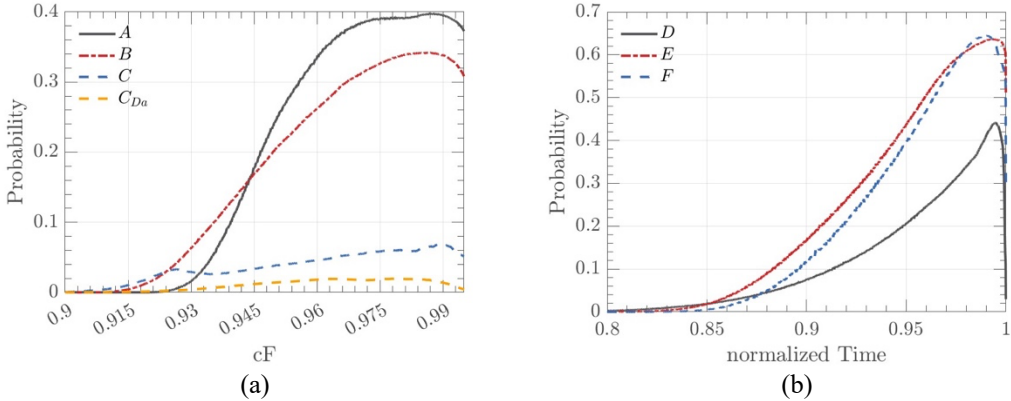

Fig. 4. Time-averaged conditioned probabilities of $T(\mathbf{x}, t) > T_{ad}$ in (a) lean flames A-C and $C_{Da}$ and (b) very lean flames D-F.

The discussed mitigation effect was already documented (using other methods) in earlier DNS [16-18,26,27] and experimental [28,29] studies. However, the present analysis, and Fig. 1 in particular, show that this effect does not indicate mitigation of all differential diffusion phenomena by turbulence. On the contrary, in the analyzed cases, an increase in $Ka$ results not only in (i) reducing (i.a) the magnitude of superadiabatic temperature, see Fig. 3, and (i.b) the probability a finding $T(\mathbf{x}, t) > T_{ad}$, see Fig. 4, but also in (ii) increasing both FCR and HRR within flame reaction zone, see Fig. 1, as well as increasing the augmentation of turbulent burning velocity due to differential diffusion phenomena, see the last column in Table 1. Such opposite effects of turbulence on superadiabatic temperature and $U_T$, FCR, or HRR is a new finding to the best of the present authors knowledge.

As hypothesized in Sect. 2, the opposite effects (specifically, the decrease in the magnitude of superadiabatic temperature and in the probability of finding it) could be associated with intensification of local mixing by small-scale turbulent eddies within flame oxidation zones. Such an interpretation is consistent with Fig. 4, which shows a decrease in the probability of finding superadiabatic temperatures with increasing $c_F$ at $c_F > c^* \approx 1$. Indeed, in a hypothetical no-mixing case, (i) $T(c_F)$ is a monotonously growing function due to termolecular radical recombination reactions with weak heat release and, hence, (ii) probability of $T(c_F) > T_{ad}$ cannot be smaller at larger $c_F$. Therefore, the reported decrease in probability of $T(c_F) > T_{ad}$ with increasing $c_F$ at $c_F > c^*$ is attributed to an important role played by mixing at $c^* < c_F < 1$, i.e., in local flame oxidation zones. Specifically, mixing of hot spots (originating from differential diffusion phenomena in the upstream local flame zones) with surrounding products, which are weaker affected by differential diffusion and, hence, are characterized by a lower temperature, results in decreasing the local temperature in the hot spots. Consequently, the probability of $T(c_F) > T_{ad}$ is decreased with increasing $c_F$ at $c_F > c^*$. Note that (i) $c^*$ varies from 0.9842 (case $C_{Da}$) to 0.9948 (case D) and (ii) the bin width is equal to 0.0001. Therefore, appearance of peak probabilities at $c_F = c^* < 1$ is a well-resolved phenomenon.

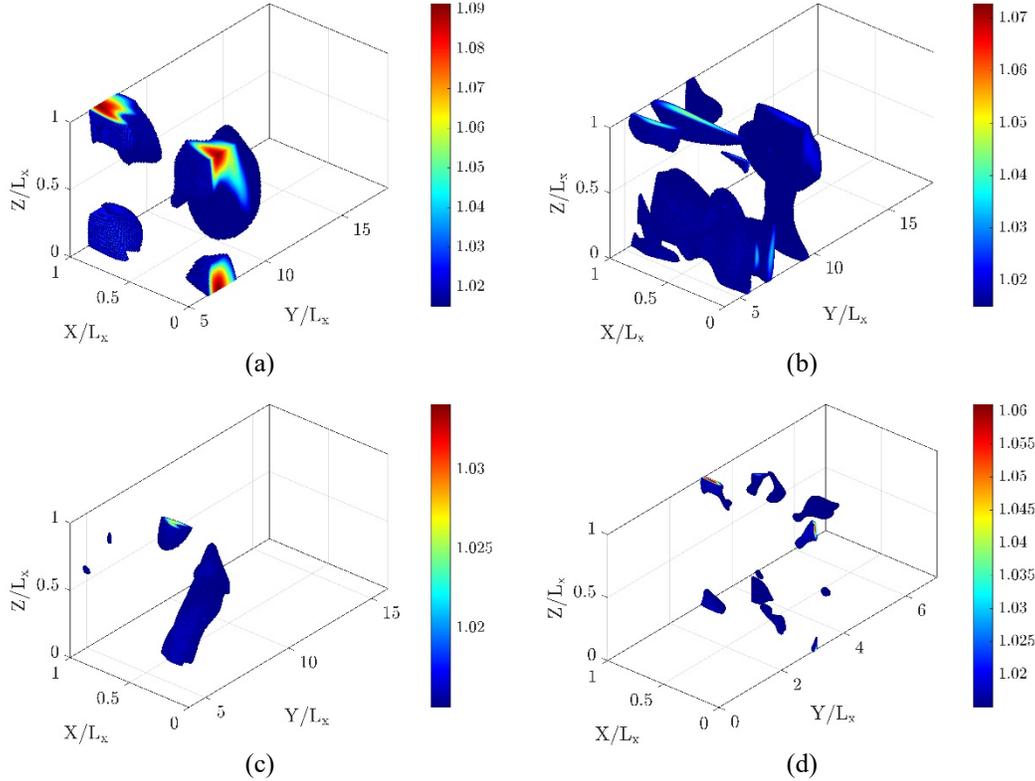

Fig. 5. Typical instantaneous images of hot spots, i.e., volumes where $T(\mathbf{x}, t) > 1.015 T_{ad}$, in lean ($\phi = 0.5$) flames (a) A, (b) B, (c) C, and (d) $C_{Da}$. Color scales show $T/T_{ad}$.

Figures 5 and 6 further support the discussed hypothesis by presenting typical images of hot spots, i.e., volumes where $T(\mathbf{x}, t) > 1.015 T_{ad}$ (the threshold 1.015 was selected to improve the image clarity). First, at lower $u'/S_L^0$, hot-spot volumes are significantly larger, cf. (i) Figs. 5a and 5b with Figs. 5c and, especially, 5d or (ii) Figs. 6b and 6c with Fig. 6a. This observation is consistent with the emphasized smoothing effect of turbulent mixing on hot spots in local flame oxidation zones[1]. Second, in line with the smoothing effect of turbulent mixing, hot spots disappear at large distances $Y/L_x$ from the mean flame front. Third, Figs.

---

[1]Hot spots cannot appear in combustion products due to the lack of chemical reactions with notable heat release there. Hot spots can only appear in local flame zones and can be convected to the products from these zones.

6b and 6c show intermittency of layers where $T(\mathbf{x},t) > 1.015 T_{ad}$ and $T(\mathbf{x},t) < 1.015 T_{ad}$, with (i) the former layers occupying entire cross-sections of the computational domain, (ii) temperature distribution in such a single layer being more uniform at larger $Y/L_x$, and (iii) temperature magnitude being smaller at larger $Y/L_x$. All these features are also consistent with the hypothesis that hot spots are smoothed out due to turbulent mixing at $c^* < c_F \leq 1$. For instance, the superadiabatic-temperature-layers become more uniform with increasing $Y/L_x$ due to turbulent mixing in transverse directions. It is also worth noting that (i) the aforementioned layers characterized by a lower $T(\mathbf{x},t)$ appear to originate at instants associated with weaker differential diffusion effects and (ii) hot spot volumes and peak $T/T_{ad}$ are significantly larger at lower $\phi = 0.35$ even if these leaner flames are characterized by higher $u'/S_L^0$ and $Ka$, cf. Figs. 6b and/or 6c with Figs. 5a and/or 5c. In other words, differential diffusion effects are greatly amplified with decreasing $\phi$.

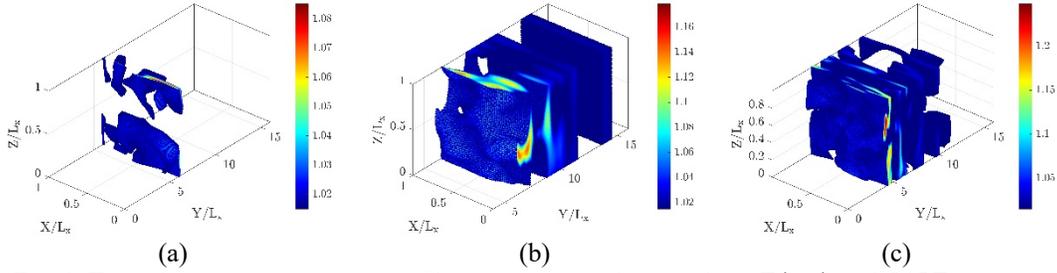

(a)  (b)  (c)

Fig. 6. Typical instantaneous images of hot spots, i.e., volumes where $T(\mathbf{x},t) > 1.015 T_{ad}$, in very lean ($\phi = 0.35$) flames (a) D, (b), and (c) F. Color scales show $T/T_{ad}$.

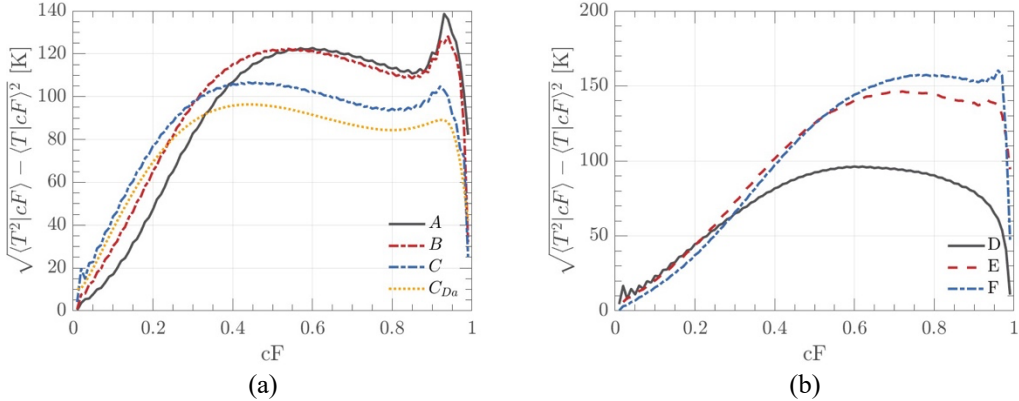

(a)  (b)

Fig. 7. Profiles of time-averaged rms conditioned temperature $\sqrt{\langle T^2|c_F\rangle - \langle T|c_F\rangle^2}$ sampled from (a) lean ($\phi = 0.5$) flames A-C and $C_{Da}$ or (b) very lean ($\phi = 0.35$) flames D-F, respectively. Profiles are sampled from the entire flame brushes and are averaged over time.

Finally, the discussed hypothesis is also in line with profiles of dimensional conditioned rms temperature, reported in Fig. 7. It shows that, at both equivalence ratios and $c_F > 0.6$, these rms values are decreased with increasing $Ka$, thus, indicating more uniform, i.e., better mixed, temperature fields. This trend is associated with intensification of turbulent mixing at higher $Ka$. Nevertheless, at $c_F < 0.9$, this mixing intensification does not result in mitigating the influence of differential diffusion on the temperature field with increasing $Ka$, because neither Fig. 2 nor Fig. 3 shows such a mitigation. Under conditions of the present study, the mixing intensification plays a major role at $c_F > 0.9$ only, i.e., in local oxidation layers.

## 5. Conclusions

The performed analyses of DNS data obtained from seven complex-chemistry lean hydrogen-air turbulent flames shows that, at sufficiently high $Ka > 1$, (i) both magnitude of superadiabatic temperature and probability of finding it are decreased with increasing Karlovitz number, but (ii) significant influence of differential diffusion effects on local structure of flame reaction zones and bulk burning rate is well pronounced in all cases, even at $Ka$ as high as 1315. Therefore, a decrease in magnitude of superadiabatic local temperature with increasing Karlovitz number or even negligible probability of finding such a high temperature at high $Ka$ is not an evidence that differential diffusion effects play a minor role. Under conditions of the present study, mitigation of phenomenon of superadiabatic temperature at high $Ka$ is attributed to intensification of turbulent mixing in local flame oxidation zones, rather than weakening differential diffusion effects in local flame reaction zones.

**Declaration of competing interest**

The author declares that they have no known competing financial interests or personal relationships that could have appeared to influence the work reported in this paper.

**Acknowledgements**

AL gratefully acknowledges the financial support by Swedish Research Council (Grant No. 2023-04407) and Chalmers Area of Advance "Transport" (Grant No. 2021-0040). Other authors have been supported in part by NSFC (Grants No. 12225204, No. 51976088, and No. 92041001), the Shenzhen Science and Technology Program (Grants No. KQTD20180411143441009 and No. JCYJ20210324104802005), Department of Science and Technology of Guangdong Province (Grants No. 2019B21203001 and No. 2020B1212030001), National Science and Technology Major Project (Grants No. 539 J2019-II-0006-0026 and No. J2019-II-0013-0033), and the Center for Computational Science and Engineering of Southern University of Science and Technology.